\documentclass[sigconf, nonacm]{acmart}

\usepackage{cleveref}
\usepackage{amsmath}
\usepackage{adjustbox}
\usepackage{multirow}
\usepackage{multicol}
\usepackage{tabularray}
\usepackage{hyperref}

\setcopyright{none}
\renewcommand\footnotetextcopyrightpermission[1]{}
\begin{document}

\title{Domain Adaptation of Multilingual Semantic Search - Literature Review}

\author{Anna Bringmann}
\email{anna.bringmann@stud.uni-goettingen.de}
\affiliation{%
  \institution{Georg-August University Göttingen}
  \country{Germany}
}
\author{Anastasia Zhukova}
\email{zhukova@gipplab.org}
\affiliation{%
  \institution{Georg-August University Göttingen}
  \country{Germany}
}
\begin{abstract}
    This literature review gives an overview of current approaches to perform domain adaptation in a low-resource and approaches to perform multilingual semantic search in a low-resource setting. We developed a new typology to cluster domain adaptation approaches based on the part of dense textual information retrieval systems, which they adapt, focusing on how to combine them efficiently. We also explore the possibilities of combining multilingual semantic search with domain adaptation approaches for dense retrievers in a low-resource setting.
\end{abstract}


\settopmatter{printfolios=true}
\maketitle

\section{Introduction} \label{introduction}
Text information retrieval (tIR) describes acquiring the relevant textual information from all available textual information, given a user-entered natural language query. The query and the text are a word sequence divided into several tokens from a specific language \cite{zhao2022dense}.   

Traditionally, tIR was performed using lexical methods. These suffer from the lexical gap since they do not include semantic information and, thus, cannot recognise synonyms or distinguish ambiguous words. Nowadays, to close the gap, natural language processing (NLP) is widely applied to tIR tasks. The most common models are dense retrievers, and their invention resulted in the concept of semantic search. Dense retrieval models map the user query and the available textual information to a shared dense vector space. The generated representations include semantic information and are closer to one another in the shared vector space if they are semantically similar. Thus, dense retrievers utilise approximate nearest neighbour search to fetch relevant information. However, large-scale labelled data is needed to train dense retrievers, and they are sensitive to domain shifts \cite{thakur2021}. 

We define a domain as a coherent type of text corpus, i.e. the specific dataset used for training. The type is a variety of latent factors, e.g. topic (chemistry vs. sports), genre (social media vs. news article) or style (formal vs. informal) \cite{ramponi-plank-2020-neural}. For most specialised domains, the availability of labelled data is scarce and collecting it is resource-intensive. Doing that manually for every language and every domain is not feasible \cite{shen2022lowresource}. Thus, automated processes for domain adaptation in low-resource setups are required and have been subject to extensive research in the past years \cite{shen2022lowresource, ren2023thorough, zhao2022dense}. Nonetheless, most domain-adaptation research focuses on monolingual dense retrieval, while the field of multilingual semantic search is rapidly advancing at the same time, holding the promise of breaking down language barriers and expanding information access across diverse linguistic contexts \cite{Yang2019, litschko2021crosslingual, Zhang2022, feng-etal-2022-language, reimers-gurevych-2020-making, Bonifacio2021}. 

However, the possibility of combining domain-adaptation techniques with multilingual approaches to semantic search has yet to be explored. Therefore, this literature review introduces a typology to explore the possibility of utilising domain-adaptation approaches in a multilingual setup. Our research question: "How to perform domain adaptation for multilingual semantic search?" formalises this objective.

The rest of this paper is structured as follows: In \Cref{method}, we introduce the methodology used to retrieve relevant literature. Afterwards, in \Cref{related}, we present related work to give an overview of the research field, exploring the possibilities of adapting multilingual semantic search to specialised domains. That is followed by \Cref{domain_adapt} in which we review existing approaches for adapting dense tIR systems to specialised domains in a low-resource setting. Subsequently, in \Cref{multilingual}, we examine existing approaches to performing multilingual semantic search. In \Cref{discussion}, we summarise the strengths and weaknesses of the different approaches and discuss their compatibility. Then, we finalise our work by giving an outlook and drawing a conclusion in \Cref{outlook}.

\section{Method} \label{method}
We applied a process consisting of four steps to find literature relevant to our research question. The first step was a keyword search, from which we selected papers based on different criteria. The keywords we used to describe semantic search were "semantic search" and "dense retrieval." We then combined these with either the keywords to describe domain adaptation in low resources settings ("domain adaptation", "low resource", "few-shot", and "zero-shot") or the keywords to describe multilingual semantic search ("multilingual", "language agnostic" and "cross-lingual"). Thus, exemplary keyword searches conducted by us were "multilingual semantic search" or "low resource dense retrieval".\footnote{applying dense retrieval to languages for which available data is scarce is also considered a low-resource setting. Therefore, this keyword combination also yields results for adapting semantic search to different languages.}

The criteria for selecting relevant papers from our initial keyword search(es) was finding existing literature reviews. We found one recent general review of dense textual retrieval \cite{zhao2022dense}, which contained information about applying dense retrievers to low-resource and multilingual settings. We selected this as an initial paper. Then, we continued with the two remaining steps. 

The second step was thoroughly reading the selected papers to decide whether they were relevant to our research question and whether we should keep them in our collection. If we considered a paper relevant, the third and fourth steps followed. The third step was reviewing the paper's references and adding the papers we missed in our keyword search to our collection of relevant papers. 

The fourth step was searching for the relevant paper on \href{https://www.connectedpapers.com/}{connected papers}, retrieving the papers connected to the paper we initially deemed relevant but published after it. The reasoning behind only retrieving the papers published after the paper we initially considered relevant was that relevant papers published before are already part of the initial paper's references, and we thus retrieved them in the third step. Using connected papers allowed us to discover recent developments in the field, which the third step could not bring forth. 

Then, we read the selected papers' abstracts, deciding which were relevant, then read the relevant ones thoroughly and performed the third and fourth steps again until nearly no new papers could be retrieved. 

Afterwards, we conducted another keyword search, using the initial keywords, to check whether we missed any relevant, recent papers. Then, we performed steps two to four for the new papers. With 250 papers in our collection of relevant papers, we stopped retrieving new literature. 

To make this literature review comprehensive, independent of readers' knowledge levels, we included a short history of performing textual information retrieval in \Cref{history}. We also recap the architecture of information retrieval systems and dense retrievers in \Cref{architectures} and explain the training procedure of bi-encoders for first-stage retrieval, including the different options for negative selection.

\section{Related Work} \label{related}
A growing body of research describes dense information retrieval systems \cite{karpukhin-etal-2020-dense, zhao2022dense, Luan2021} and the topic of adapting dense retrievers to new domains in low-resource settings gained attention within the last years \cite{Li2022, wang2022, xu2022laprador, lee2021learning}. Some literature reviews summarising state-of-the-art approaches for domain-adaptation of dense retrievers exist \cite{ren2023thorough, shen2022lowresource}. For example, Ren et al. \cite{ren2023thorough} analysed the influence of training data properties on the zero-shot performance of dense retrievers and compared different existing methods for domain adaptation in zero-shot settings. 
The same is true for multilingual semantic search \cite{Mao2022EMSEA, feng-etal-2022-language, reimers-gurevych-2020-making, Yang2019, litschko2021crosslingual, Cer2017, Zhang2022, asai-etal-2021-xor}.
While \cite{hedderich2021survey} considers multilingual dense retrieval and adapting dense retrievers to domains with specialised language in low-resource setups, they do not investigate the possibility of combining both. 
Our focus on the possibility of adapting multilingual dense retrievers to another low-resource domain is, therefore, a new perspective on existing research. 

To explain how to enable multilingual dense retrieval in specialised domains, we first need to look at the structure of information retrieval systems and the model architectures used for their different parts since the different approaches for domain adaptation of multilingual semantic search adapt different parts of that architecture. We will do so in the following section.  

\section{Information Retrieval systems} \label{architectures}
We formalise the Aim of text IR systems as: \newline
Given the natural language query $q$ and a collection of $m$ documents $D =$ {$d_i$}$^m_{i=1}$ the IR system returns a list $L = [d_2, d_2, ...., d_n]$ of $n$ relevant documents sorted by the retrieval model's relevance scores. 

\begin{figure}[H]
    \centering
    \includegraphics[width=0.48\textwidth]{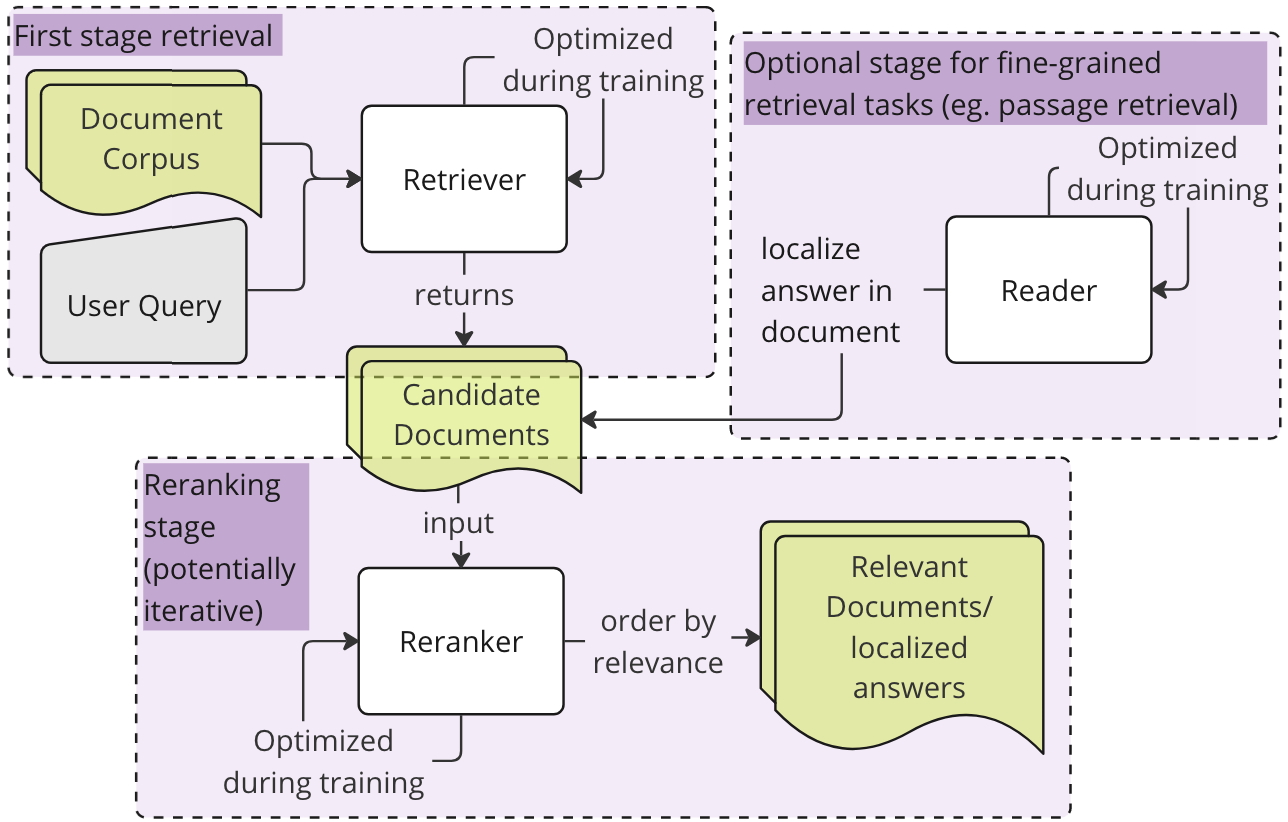}
    \caption{Information Retrieval Systems}
    \label{fig:IR_system}
\end{figure}

To calculate these relevance scores, sparse information retrieval models perform lexical matching, while dense retrievers perform semantic matching. All information retrieval systems consist of pipelines that combine multiple steps as visualised in \Cref{fig:IR_system}. 

The first stage performs the retrieval. Thus, the model selects several candidate documents relevant to the user query. The model which implements this stage is called the retriever. During the second, the re-ranking stage, the selected candidates are ordered by importance. The model which implements this stage is called the reranker. TIR system can utilise one to many re-ranking stages to refine the results. A model called the reader performs the third stage of textual information retrieval. We only add it if the information retrieval task is more fine-grained than document-level retrieval. The reader analyses the documents the retriever returned and localises the response to the query \cite{zhao2022dense}. 

\subsection{Dense Retrievers}
Based on the formalisation of tIR systems, a way of formalising dense retrievers arises: \newline
Dense retrieval models encode the query and the document corpus as dense vectors. Thus, they compute relevance scores through some similarity function between these dense vectors: 
\begin{equation}
    \text{Rel}(q,d) = f_{sim}(\phi(q), \psi(d))
\end{equation}
where $\phi(\cdot) \in \mathbb{R}^l$ und $\psi(\cdot) \in \mathbb{R}^l$ are functions that map the query and the documents into $l$-dimensional vector space. A deep neural network performs this mapping, and the similarity $f_{sim}$ is measured by, for example, the inner product or the cosine similarity. Dense retrievers, thus, measure the semantic interaction of query and text based on the learned representation of both in latent semantic space. The closer the query and text are to one another in latent space, the more similar they are \cite{zhao2022dense}. Apart from this formal notation, dense retrievers can be divided into two mainstream architecture types: cross- and bi-encoders. We introduce both in the following subsection.

\subsection{Bi-encoders, Cross-encoders or Both?} 
\begin{figure}[H]
    \centering
    \includegraphics[width=0.48\textwidth]{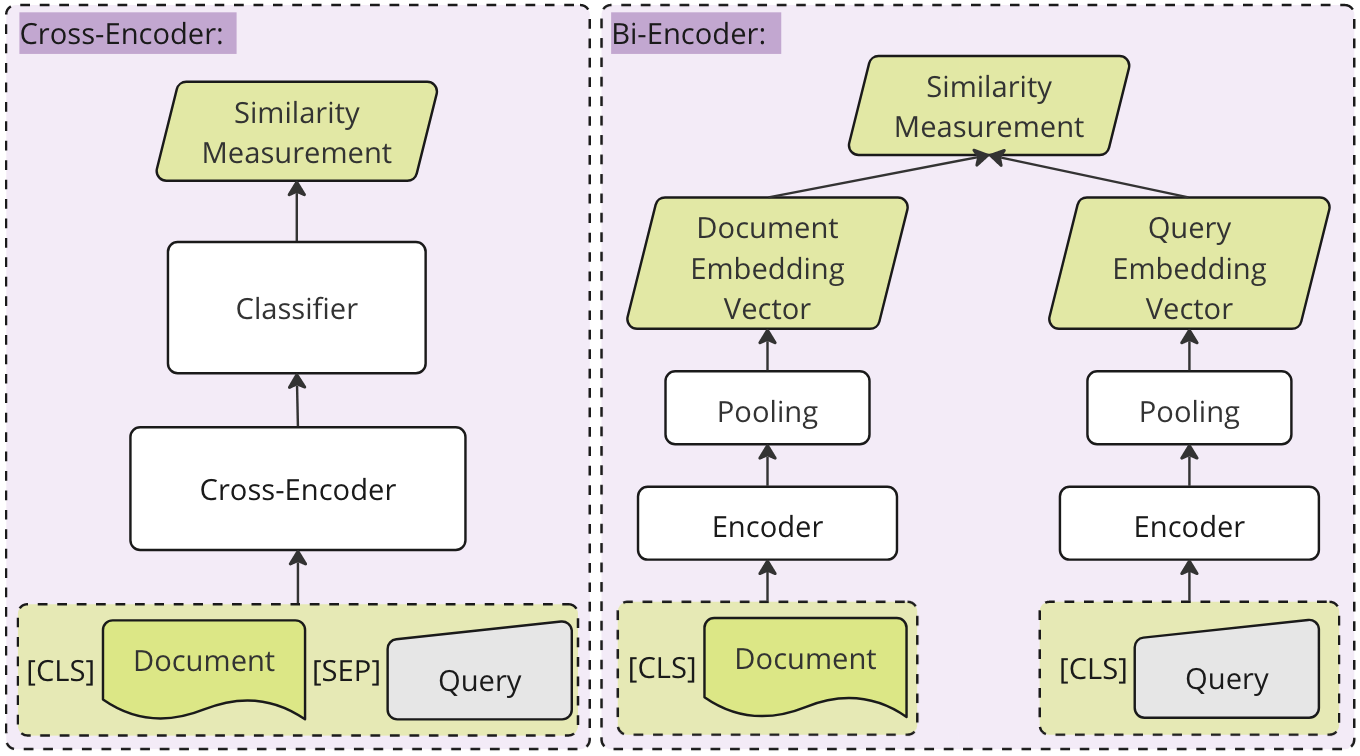}
    \caption{Dense Retriever Architecture Types}
    \label{fig:dr_architecture}
\end{figure}

First, we introduce bi-encoders and, afterwards, cross-encoders. However, \Cref{fig:dr_architecture} allows a direct, visual comparison of both architecture types. We finalise this section by analysing the advantages and disadvantages of both architectures.

As shown in \Cref{fig:dr_architecture}, bi-encoders are based on a two-tower architecture, utilising PLM-based encoders. They perform the self-attention mechanism for query and document separately, mapping each to the same dense vector space and then utilising a similarity measurement to measure the distance between both \cite{wang2022, zhao2022dense, Huang2013}. Their two-tower architecture makes bi-encoders flexible since document and query encoder architecture can differ. It resulted in two different bi-encoder structures. 
The first is called single-representation bi-encoder. It embeds the query and text separately using two different pre-trained language models \cite{karpukhin-etal-2020-dense, qu-etal-2021-rocketqa, xiong2020approximate}. We place the special [CLS]  token at the beginning of the query, and the text and their encodings represent their semantic meaning. The relevance score for a query text combination is then computed through some similarity function using the [CLS] token embeddings. The downside of the single-representation bi-encoder structure is that it does not accurately capture the semantic information of query and text. The second bi-encoder structure is called multi-representation bi-encoder. It aims at enhancing the semantic information stored in the embeddings by combining multiple query and document embeddings to measure their similarity from different "semantic viewpoints". These contextualised embeddings are learned and stored during training as well as indexing. This second bi-encoder structure improves the retrieval performance compared to the single-representation structure. However, storing the multiple generated views makes their index large, leading to higher computational and memory costs. Examples for multi-representation bi-encoders are poly-encoder \cite{humeau2020polyencoders}, ME-BERT \cite{Luan2021}, ColBERT \cite{khattab2020colbert}, ColBERTer \cite{hofstätter2022introducing}, MVR \cite{zhang2022multiview} and MADRM \cite{Kong2022}. 

In contrast to bi-encoders, cross-encoders receive the concatenated query and text pair, distinguished by a separation token [SEP], as input as shown in \Cref{fig:dr_architecture}\cite{qiao2019understanding}. They utilise a cross-attention mechanism to compute the interaction between any two tokens in the input and encode every token in vector space. Thus, cross-encoders enable the tokens to interact across queries and text \cite{GuoetAl2017, Xiong_2017}. We can then compute match representations for the query-text pair using the learned representations. Primarily, the encoding of the [CLS] token is used for this semantic matching \cite{zhao2022dense, wang2022}. Nevertheless, we can also average over all token embeddings to calculate the similarity measurement \cite{Reimers2019}. The cross-encoder's output is a fine-grained relevance score for the query text pair, and a higher score indicates higher relevance of the text, given the query.
 
Utilising a bi-encoder or a cross-encoder has different up- and down-sides. Their two-tower architecture makes bi-encoders computationally more efficient than cross-encoders since approximate nearest neighbour search enables fast recall of large-scale vectors, and only the query, but not the textual information, must be encoded during query time. Since cross-encoders calculate relevance scores for every possible query and text pair, end-to-end information retrieval is not possible, as they do not create independent representations of query and text. Thus, the encoding must be recomputed every time for every query text pair \cite{zhao2022dense, wang2022}. However, bi-encoders need more training data than cross-encoders since they independently map the documents to vector space. They also reach lower retrieval accuracy. 

Due to these pros and cons, bi- and cross-encoders are suited for different tasks in information retrieval systems \cite{zhao2022dense}. Bi-encoders are used in first-stage retrieval to fetch the candidate documents, while Cross-encoders are adapted as re-rankers or readers. In general, cross-encoders perform better if applied to zero-shot retrieval tasks on out-of-domain data, and multiple methods of domain adaptation dense retriever in low-resource setups rely on cross-encoder architectures \cite{wang2022}. However, some tricks can be applied during bi-encoder training for first-stage retrieval to improve their ability to generalise to new domains in a low-resource setup. Therefore, the following section will give an overview of how to train bi-encoders.

\subsection{Training bi-encoders for first stage dense retrieval}
First, we outline how to train a first-stage dense retriever with a bi-encoder architecture. Afterwards, we summarise how to select negative documents for that setting.

We jointly optimise a query and a document-oriented loss function while training first-stage retrievers.

The query-oriented loss function is the 
exact negative log-likelihood \cite{karpukhin-etal-2020-dense}. It maximises the probability of the relevant (positive) document with respect to the query. The formula for this positive query-document pair is: 
\begin{equation}
L(q_i, d_i^+) = -log \frac{e^{f(\phi(q_i), \psi(d_i^+))}}{e^{f(\phi(q_i), \psi(d_i^+))}+\sum_{d'\in D^-}e^{f(\phi(q_i), \psi(d'))}}
\end{equation}
where $d_i^+$ is a relevant document for query $q_i$, $\phi(\cdot)$ and $\psi(\cdot)$ are the query and text encoder and $f(\cdot)$ measures the similarity between query embedding $\phi(q_i)$ as well as text embedding $\psi(d_i^+)$. $D^-$ is the set of all documents except the positive one(s) 
Since the formula iterates over all indexed documents in the normalisation term computing, it is very time-consuming. The negative sampling trick was introduced to the negative log-likelihood to tackle that issue \cite{karpukhin-etal-2020-dense, qu-etal-2021-rocketqa}. It reduces computational costs compared to the exact formulation by sampling a set of negatives from all documents. Its objective can be summarised as increasing the likelihood of positive documents while decreasing the likelihood of sampled negative documents. We formalise it as:
\begin{equation}
    L(q_i, d_i^+) = -log \frac{e^{f(\phi(q_i), \psi(d_i^+))}}{e^{f(\phi(q_i), \psi(d_i^+))}+\sum_{d'\in N_{q_i}}e^{f(\phi(q_i), \psi(d'))}}
\end{equation}
where $N_{q_i}$ is a small set of negative samples for the query $q_i$. It was derived from the InfoNCE loss \cite{Oord2018}, which contrasts a positive pair of examples with randomly sampled examples. 

The document-oriented loss function, which we jointly optimised with the query-oriented loss function to train first-stage retrievers, is the negative log-likelihood oriented at text \cite{xu2022laprador}: 
\begin{equation}
L_T(q_i, d_i^+) = -log \frac{e^{f(\psi(d_i^+), \phi(q_i))}}{e^{f(\psi(d_i^+), \phi(q_i))}+\sum_{q^-\in Q^-}e^{f(\psi(d_i^+), \phi(q^-))}}
\end{equation}
where $Q^-$ is a set of sampled negative queries. 
In the following sections, we abbreviate the combination of the query and document-oriented loss functions, including the negative sampling trick as "NLL+NS".

To make the differentiation between positive and semantically close negative query-document pairs more precise. We can add a text-text similarity constraint to the loss of the bi-encoder performing the first-stage retrieval \cite{Ren_2021}. The constraint ensures that the similarity between a positive document and the respective query is bigger than between a positive document and a negative document. This objective can be formalised as:
\begin{equation}
f_{sim}(d^+, q_i) > f_{sim}(d^+, d_-)
\end{equation}
The complete loss function with the constraint is:
\begin{equation}
L_{TT}(q_i, d_i^+) =-log\frac{e^{f(\psi(d_i^+), \phi(q_i))}}{e^{f(\psi(d_i^+), \phi(q_i))}+\sum_{d'\in N_{q_i}}e^{f(\psi(d_i^+), \psi(d'))}}
\end{equation}
The sum in the normalisation term incorporates the text-text similarity.

One of the main training issues of tIR systems arises from the NLL+NS loss. It is based on the large-scale candidate document space, while only a few are relevant to the query. Computational restrictions only allow us to sample a few negative documents for every query during training. Thus, the candidate space of documents shifts during training compared to testing for which we use the entire collection. Therefore, selecting negative documents during training is crucial to retrieval performance as it influences convergence for the gradient norms. Negative selection also impacts the capability of a trained retrieval model to adapt to out-of-domain data distributions \cite{karpukhin-etal-2020-dense, qu-etal-2021-rocketqa, xiong2020approximate}. Therefore, we will now introduce three prevalent negative selection strategies. 

The first strategy selects in-batch negatives (IB) \cite{henderson2017efficient, karpukhin-etal-2020-dense}. It uses all documents, except the positive one, per batch as negatives. The main reason to apply this strategy is that dense retrievers are often optimised on GPU in batches. Thus, we cannot access the whole document corpus during training since keeping that in memory is unfeasible. IB outperforms random sampling of negatives from the whole document corpus before training since performing the sampling per batch during training leads to more different negatives. 

However, applying the second approach for negative selection, called cross-batch negative selection, enables the first stage retriever to train on even more negatives \cite{qu-etal-2021-rocketqa, 
gao2021scaling}. We train the model in parallel on the different GPUs in multi-GPU settings. To utilise the negatives across GPUs. We first compute the text embeddings on all GPUs, then communicate those between GPUs and use them as negative documents for the query. However, cross-batch negative selection can also be used in a single GPU setting if gradient cashing is applied, accumulating mini-batches for negative selection \cite{gao2021scaling}. However, the larger candidate document space for negative selection provided by cross-batch negative selection comes at the cost of a higher computational effort than in-batch negative selection. 

While in-batch and cross-batch negatives only increase the number of negatives, the third strategy, hard negative selection, tries to increase the quality of selected negatives. Hard negatives are negative documents semantically similar to positive examples but irrelevant to the query. Three different ways of selecting hard negatives exist. The first way generates static hard negatives. Thus, during training, the negative selector is fixed. Different methods can be used for the selection of static hard negatives. We can use randomly selected top results from another sparse or dense retriever, for example, the BM25 results, which are lexically similar but not relevant to the query \cite{karpukhin-etal-2020-dense}). Another option is to mix multiple types of negatives. For example, retrieved negatives from BM25 and heuristic-based context negatives \cite{lu2020neural}. The combination performed in the second method leads to better retrieval performance than the first. We can also apply different fusion strategies for differently generated hard negatives to improve retrieval performance even further \cite{lu-etal-2021-multi}. Hofstätter et al. \cite{hofstätter2021efficiently} proposed topic-aware sampling of hard negatives, an in-batch sampling technique that implicitly generates hard negatives. They first cluster the queries, leading to higher similarity of samples in a batch, and then sample the negative documents from one cluster for every batch. 

The second way to select hard negatives is to update them periodically during training. Since the training of dense retrievers happens iteratively, updating the hard negatives alongside the iterations is more suitable than fixing them. Periodically updating the hard negatives generates more informative negatives for training the discriminator, and they allow to weaken the train-test discrepancy in the document candidate space since more informative negatives are "seen" during training \cite{zhao2022dense}. However, dense retrievers must be warmed up before training with dynamic hard negatives, increasing the computational costs associated with this method \cite{xiong2020approximate}. Different strategies for utilising dynamic hard negatives have been explored. ANCE \cite{xiong2020approximate} samples hard negatives from the top results returned by the in-training retriever itself (global hard negatives). The advantage of this approach is fast convergence. The disadvantage is that indexed document embeddings must be updated after updating the model parameters, which is time-consuming. A solution is to use an asynchronous index refresh strategy that periodically updates the index for each $m$ batches. Another strategy is called ADORE \cite{xiong2020approximate}. It samples documents from retrieval results based on the in-training query encoder. The text encoder and text embedding index are fixed. Since the query encoder for negative selection is optimised during training, adaptive negatives for the same query are generated. 

The third way of selecting hard negatives uses denoising strategies to ensure that the selected hard negatives contain fewer false negatives. Again, multiple methods applying this third strategy exist. RocketQA \cite{qu-etal-2021-rocketqa} uses a cross-encoder to sample the top documents likely to be false negatives. They use the hard negatives for the bi-encoder as candidate documents for the cross-encoder to enable faster computation, thus refining the hard negatives selected to train the bi-encoder. SimANS \cite{zhou2022simans} introduces another method. The authors use ambiguous negatives. Thus, negatives ranked close to positives, with moderate similarity. They prove these are less likely to be false negatives and more informative. 

Empirical studies showed that in-batch negatives are not informative enough to train first-stage dense retrievers \cite{xiong2020approximate, lu-etal-2021-multi, lu2020neural}. Since the method only considers a small subspace of candidate documents as possible negatives during training, it does not sample from the entire distribution. This smaller candidate space results in the probability of sampling informative negatives being close to 0 \cite{xiong2020approximate}. However, according to Zhan et al. \cite{zhan2021optimizing}, hard negative and random negative sampling optimise different retrieval objectives. Random negative sampling minimises the total pairwise error, possibly overemphasising learning to handle complex queries. In contrast, hard negative sampling minimises the top pairwise error, which is more suitable for optimising top rankings. Combining both strategies during training is a valuable approach already used to adapt dense retrievers to new domains, which we discuss in the next section.

\section {Domain Adaptation of Semantic Search} \label{domain_adapt}
Multiple strategies have been devised to facilitate the integration of dense tIR systems into new domains with limited resources. To comprehensively explore these approaches, we categorised them based on the components of the IR system they modify. This categorisation is particularly valuable, as the potential for combining distinct strategies becomes more feasible when they target different aspects of the IR pipeline.

The approaches that adapt the data for training a tIR system either generate (pseudo) query document pairs - utilising query generation or contrastive learning - or enhance the quality of available data through semi-supervised knowledge distillation. Other approaches directly alter the dense retrievers by scaling up the model size or utilising a more capable model. The approaches that adapt the training of models used for tIR employ a multi-task learning approach, aim to create a domain-invariant embedding space or utilise parameter-efficient learning to use pre-training techniques efficiently. An approach that directly modifies the document ranking output by tIR systems integrates sparse retrieval to enhance the system's term-matching capability. In the following, we will discuss these approaches in detail and close this section by analysing their compatibility.

\subsection{Data Adaptation}
The main problem of domain adapting dense tIR systems is that, often, no labelled data for the target domain is available. However, unlabelled documents are easier to obtain. Thus, generating query-document pairs from unlabelled text data is one approach to the domain adaptation of dense IR systems. Forming these pairs can be done by query generation or contrastive learning.

\subsubsection{Query Generation}

Query generation approaches form positive query-document pairs by generating queries not found in the documents, utilising a data generation model \cite{Liang2020, wang2022, reddy2021robust, ma2021zeroshot, Dai2022Promptagator}.
The query generation is achieved by combining a Query Generator (QG) with a filtering mechanism to enhance the quality of the generated queries. Three different categories of QGs can be distinguished. Rule-based QGs apply handcrafted features and templates to generate queries. However, they are time-consuming to design, domain-specific and only generate certain question forms \cite{Pandey2013AutomaticQG, Rakangor2015LiteratureRO}. 
Prompt-based QGs utilise generative PLMs, which can be presented with documents and the prompt to generate a query \cite{bonifacio2022inpars, sachan2023improving, dai2022dialog, Dai2022Promptagator}. Supervised QG use a pre-trained language model, fine-tuned on in- or out-of-domain data. 

These supervised GQs can be trained end-to-end. Then, they directly generate query-text pairs as training data for the dense retriever. Shakeri et al. \cite{shakeri-etal-2020-end} used this approach. They utilised a pre-trained language model and reframed the task of question-answer generation to machine reading comprehension. Based on that, they trained a seq2seq network that generates a question-answer pair given an input text. Reddy et al. \cite{reddy2021robust} adapted this idea and added a selection step to the approach, enabling them to generate better question-answer pairs. 

The second approach to training supervised QGs utilises a pipeline consisting of multiple dense retrievers and generative models. Strategies that employ this approach are QGen \cite{ma2021zeroshot}, which uses a query generator trained on general domain data to generate domain queries for the target corpus. Then, a dense retrieval model is trained on this (pseudo) labelled data from scratch. GPL \cite{wang2022} extends the QGen approach by using (pseudo) labels from a cross-encoder combined with hard negatives. Alberti et al. \cite{alberti2019synthetic} used a large text corpus to construct question-answer-pairs in three stages: First answer extraction, second question generation and third roundtrip filtering. Lewis et al. \cite{lewis2021paq} extended this approach by adding passage selection and global filtering. The supervised QGs generate the best queries, while the other GQs often generate inadequate results \cite{shen2022lowresource}. 

The query generators can handle different inputs. If only a document is passed, the QG can attend to different document spans to generate a query and thus even generate multiple queries \cite{du-cardie-2017-identifying, kumar-etal-2019-cross}. Alternatively, an answer span from the document and the document can be passed to the QG, generating a query conditioned on both \cite{alberti2019synthetic, shakeri-etal-2020-end}. Some QGs can also handle more fine-grained information like question types to generate \cite{cao-wang-2021-controllable, gao-etal-2022-makes}. Other QGs can process further context to make queries less ambiguous, reducing the question entropy, thereby making learning easier for the QG while increasing the probability of error propagation \cite{liu2022collaborative, zhang2019addressing}. 

The second part of the query generation approach is a filtering mechanism to enhance the quality of the generated queries. Different choices of filters were explored. Round trip consistency can filter out low-quality generated queries \cite{alberti2019synthetic, dong2019unified}. It uses a pre-trained QA system to generate an answer based on the generated queries. The query is only kept if the generated answer is the same as the answer used to generate the query. An alternative filtering mechanism manually sets an acceptance threshold to relax this rather strict condition. Possible metrics for thresholding are the probability score provided by the QA system which generated the answer from the (pseudo) query \cite{zhang2019addressing, lewis2021paq}, the LM score from the QG generator itself \cite{shakeri-etal-2020-end, Liang2020}, or an entailment score which can be acquired from a model trained on question-context-answer pairs \cite{liu2022collaborative}. Other filtering mechanisms are influence functions to estimate the effect of including the synthetic query on the validation loss \cite{yang-etal-2020-generative}, ensemble consistency for which an ensemble of QA models are trained and only the questions onto which most QA systems agree are kept \cite{bartolo-etal-2021-improving}. If some in-domain labelled samples are available, possible filter mechanisms are reweighting the (pseudo) samples based on the validation loss \cite{sun2021fewshot} or selecting samples that lead to validation performance gains \cite{yue2022synthetic}.

\subsubsection{Contrastive Learning}

However, training another model to generate query document pairs is computationally expensive. Self-contrastive learning presents another avenue for constructing (pseudo) query-document pairs. It involves utilising various parts of documents or external knowledge as (pseudo) queries. Different strategies have been proposed for constructing positive query document pairs.

Perturbation-based methods introduce augmentations to the original text, treating the augmented and original text as positive pairs \cite{shen2022lowresource}. Possible augmentation methods are word deletion, substitution or permutation \cite{Zhu_2021, yu2022cocolm}, adding drop out in the representation layers \cite{Gao2021} or passing sentences through multiple language models \cite{carlsson2021semantic}. 

Summary-based methods for constructing positive pairs involve extracting document segments that likely summarise the content and using them as (pseudo) queries. Suitable document parts are the document title \cite{macavaney_neuir2017-nyt, MacAvaney_2019, mass-roitman-2020-ad}, a random sentence from the first document section \cite{chang2020pretraining}, randomly sampled n-grams from the document \cite{chang2020pretraining} or generating several keywords from the document \cite{Ma_2021}. 

Proximity-based methods form positive pairs based on proximity positions in documents. Different approaches are ICT \cite{lee2019latent}, which uses a random sentence from a passage as the query and the same passage, but without the selected sentence, as the positive document. ICT can be combined with noise, for example, drop-out masks \cite{xu2022laprador} or word chopping/ deletion \cite{izacard2022unsupervised} to improve robustness. Other proximity-based approaches are selecting random spans from the same document as positive pairs \cite{gao2021unsupervised, Ma_2022}, selecting sentences from the same paragraph and paragraphs from the same document as positive pairs \cite{di-liello-etal-2022-pre}. 

Co-occurrence-based methods utilise sentences containing a co-occurrent span for constructing positive query document pairs \cite{ram2021fewshot}. One can, for example, use random sentences from the document corpus to construct a (pseudo) query, using a term in the sentence as the answer and replacing it with a unique token. Then, using the answer term to retrieve passages with BM25, these passages are the (pseudo) positive documents \cite{glass-etal-2020-span}. Alternatively, one can use a token span and its surrounding context to form a (pseudo) query and a passage that contains the same span as the (pseudo) document \cite{ram2022learning}. However, using the second approach to construct the training data biases it towards sparse retrieval due to the high lexical overlay in query-document pairs. Thus, it is unsuitable for comparing dense and sparse retrieval approaches.
    
Hyperlink-based methods construct the positive query document pairs for contrastive learning by exploiting hyperlink information, often capitalising on anchor-document relations.
They assume that anchor-document relations approximate query-document relevance and use these relations to build the positive query-document pairs \cite{Zhang2020AnUS, ma2021zeroshot}. An example is to use a sentence from the first section in a document as the (pseudo) query since this section often describes or summarises the topic and a sentence from another document that contains a hyperlink to the first document as the (pseudo) positive document \cite{chang2020pretraining}. Another example is replacing one token with a w-question phrase to form the (pseudo) query. A passage containing the replaced token/entity from a document containing a hyperlink to the first document can be used as the (pseudo) relevant document \cite{yue2022cmore}. 
The hyperlink approach performs best in practice, but documents are only sometimes linked. Thus, it is not always applicable \cite{sun2021fewshot}. Utilising reinforcement learning to select suitable (pseudo) query document pairs automatically is another option instead of settling for one of the five approaches. However, it makes the training procedure even slower \cite{Zhang_2020}. Apart from generating (pseudo) query document pairs, contrastive learning can also be used as task-adaptive pre-training. Zhao et al. \cite{zhao2022dense} performed a more in-depth discussion of dense retrieval pre-training. 

\subsubsection{Knowledge Distillation} 

The two query generation methods discussed so far only need unlabelled text data as input and aim at generating (pseudo) query document pairs. The following method, knowledge distillation, requires independent documents and queries from the target domain, which must still be linked as pairs. Knowledge distillation automatically establishes the missing link, forming (pseudo) query-document-pairs and generating relevance judgments.

The general definition of knowledge distillation in machine learning is to transfer knowledge from a more capable model (teacher) to a less capable model (student). For dense retrieval, we use it to improve the bi-encoder's (student) performance by utilising a more capable model (teacher). Knowledge distillation is a form of distant supervision \cite{thakur2021augmented, chen2022salient, wang2022}. The domain adaptation of the dense tIR system is performed by using the teacher to generate (more) precise (pseudo) relevant judgments for the document. Afterwards, we use the generated (pseudo) labelled data to train the student model.

Different models can be used as teachers in a knowledge distillation setting. If a pre-trained cross-encoder is used, we can either train it independently from the student bi-encoder or incorporate information interaction of teacher and student model when training the teacher \cite{qu-etal-2021-rocketqa}. We do so by randomly sampling the top-ranked texts of the student network as negatives for training the teacher. Thus, the cross-encoder can adapt to bi-encoder results. Another approach utilises two cross-encoders as teachers and trains them using in-batch and pairwise negatives to improve the student model's retrieval performance \cite{hofstätter2021efficiently}. Both using a single teacher as well as an ensemble of teachers improves the performance of the student model \cite{hofstätter2021improving}. An enhanced bi-encoder (e.g. ColBERT \cite{khattab2020colbert}) can also be used as a teacher model \cite{lin2020distilling}. Sometimes, even using a sparse retriever as the teacher network is desirable since dense retriever (DR) and sparse retriever (SR) have complementary strengths and weaknesses. SRs are better at exact matching and retrieving long documents as well as generalising to out-of-domain distributions, while DRs can perform semantic matching \cite{Chen_2022, Luan2021, thakur2021augmented, chen2022salient}. Utilising an SR as the teacher network for the bi-encoder enhances its term-matching capabilities. The resulting model reaches a better performance than only BM25 on test data. It enables the student model to match rare entities, thus improving the zero-shot out-of-domain performance \cite{chen2022salient}. Knowledge distillation might be more efficient if the gap between the student and the teacher network is smaller \cite{rezagholizadeh-etal-2022-pro, zhou2021understanding}. Thus, progressive distillation adaption is helpful if the teacher model is way more capable than the student model \cite{zeng2022curriculum, lu2022erniesearch, lin2023prod}. It can be implemented in two ways: either gradually making the teacher model more capable at different stages of distillation to mimic the student model's learning \cite{lu2022erniesearch, lin2023prod} or fixing the strong teacher model, but gradually increasing the difficulty of the distilled knowledge \cite{zeng2022curriculum, ren2023rocketqav2}.

The outputs generated by the teacher model can be passed to the student as either hard or soft labels. Hard labels are generated by directly setting the relevance scores generated by the teacher model as binary data labels. To exclude false positives, one can use thresholding methods (e.g. all documents with a higher score than 0.9 = positives, with <0.1 = negatives, discard rest) \cite{qu-etal-2021-rocketqa}. In this approach, the finer-grained output of the teacher model is ignored, but it utilises the same training objective as standard DR training. 
Soft labels are generated by tuning the student model to approximate the teacher's output using different distillation functions. This approach allows utilising the fine-grained output generated by the teacher model.
The MSE loss minimises the distance between the relevance scores of student and teacher model:
\begin{equation}
L^{KD}_{MSE} = \frac{1}{2}\sum_{q \in Q}\sum_{d \in D}(r_{q,d}^{(t)} - r_{q,d}^{(s)})^2
\end{equation}
where $Q$ is the set of all queries and $D$ is the set of all documents and $r_{q,d}^{(t)}$ are the relevance scores of document $d$ with respect to query $q$ as assigned by the teacher model, while $r_{q,d}^{(s)}$ are the relevance scores of document $d$ with respect to query $q$ as assigned by the student model.\newline
The KL-divergence loss normalises the relevance scores of the candidate documents as probability distributions per query and then reduces their KL-divergence: 
\begin{equation}
L^{KD}_{KL} = -\sum_{q \in Q, d\in D}(log \tilde{r}_{q,d}^{(s)} - log \tilde{r}_{q,d}^{(t)})
\end{equation}
where $\tilde{r}_{q,d}^{(s)}$ is the normalised probability distribution of the candidate documents by query of the student model's relevance scores and $\tilde{r}_{q,d}^{(t)}$ is the normalised probability distribution of the candidate documents by query of the teacher model's relevance scores. \newline
The Max-Margin loss penalises the inversions of rankings generated by the retriever:
\begin{equation}
L^{KD}_{MM} = \sum_{q, d_1, d_2} \text{max}(0, \gamma-\text{sign}(\Delta^{(t)}_{q, d_1, d_2}   \Delta^{(s)}_{q, d_1, d_2})
\end{equation}
where $q \in Q$ and $d_1, d_2 \in D$, while $\Delta^{(t)}_{q, d_1, d_2} = r_{q,d_1}^{(t)} - r_{q,d_2`}^{(t)}$ and $\Delta^{(s)}_{q, d_1, d_2} = r_{q,d_1}^{(s)} - r_{q,d_2`}^{(s)}$ \newline
$\gamma$ is the margin, $sign(\cdot)$ is the sign function, which indicates whether the value is positive, negative or zero. \newline
The Margin MSE loss minimises the margin difference of a positive-negative document pair between the teacher and student model 
\begin{equation}
L^{KD}_{M-MSE} = MSE(r_{q,d^+}^{(t)} - r_{q,d^-}^{(t)}, r_{q,d^+}^{(s)} - r_{q,d^-}^{(s)})
\end{equation}
            
Izacard and Grave \cite{izacard2022distilling} find that the KL divergence loss leads to better distillation than MSE and max-margin loss for question-answering tasks but that the Margin-MSE loss is most effective compared to the other options. This result might be attributed to margin-based losses, making the model less likely to overfit the synthetic training data. Thus, using those should be preferred if the teacher's output contains significant noise.

Knowledge distillation aims to adapt the bi-encoder model to the target-domain data distribution. Its bottleneck is the quality of the teacher model.
If the teacher's output contains much noise, the student model will overfit the noise. A possible solution for that problem, if the teacher's output scores display actual output confidence, is to use confidence-based filtering as a noise-resistant algorithm \cite{mukherjee2020uncertaintyaware, yu2021finetuning}. However, the teacher's output scores are less or un-reliable in challenging tasks  \cite{zhu2023meta}. Another issue of knowledge distillation is its' computational costs \cite{xu2020computationefficient, rao2022parameterefficient}. However, if enough computational power is available, knowledge distillation can be used to generate more fine-grained relevance labels, enhancing the quality of the training data, and it can be combined with query generation methods.

All three methods augmenting the tIR system's data level have the same problem. The generated labels tend to be noisy, meaning they contain more errors than manually labelled data. Training on such noisy data can hurt the model's performance. Thus, two approaches to handle the noise have been explored \cite{hedderich2021survey}. The first is noise filtering. Hard noise filtering removes instances that are incorrectly labelled with a high probability. We can find the incorrect labels through a probability threshold \cite{jia-etal-2019-arnor}, a binary classifier \cite{Adel2018CISAT, onoe2019learning, huang-du-2019-self} or a reinforcement-based agent \cite{yang-etal-2018-distantly, nooralahzadeh-etal-2019-reinforcement}. Soft noise filter reweights the labelled query-document-pairs according to their probability of being incorrect \cite{le-titov-2019-distant} or through attention measurements \cite{hu-etal-2019-improving}. The second approach to handle the noise in the data is noise modelling. It can be performed using correlation coefficients to estimate the relationship between clean and noisy labels \cite{chen2019uncover, hedderich2021anea}. Then, a noise model is applied to shift the distribution of the noisy data to the (unseen) distribution of the clean data. Afterwards, we train the dense retrievers on the data with the new "clean" distribution. Chen et al. \cite{chen2020relabel} use multiple reinforcement agents to re-label noisy instances. Ren et al. \cite{ren-etal-2020-denoising} apply multiple sources of distant supervision and learn how to combine them.
Most of these denoising approaches were suggested for named entity recognition tasks. However, their strategies also apply to the data adaptation methods introduced in this section.

\subsection{Model Adaptation}
Apart from tackling the problem of domain-adapting dense retrievers in low-resource scenarios on the data level, we can also adapt the models themselves. 
Zhan et al. \cite{zhan2022evaluating} showed that cross-encoders are more capable of generalising to out-of-domain test data than bi-encoders and sparse retrievers. However, their high computational costs during inference make their practical application impractical. Ni et al. \cite{Ni2021LargeDE} scaled up the size of a T5 dense-retriever through multi-stage training and reached a better zero-shot retrieval performance on several BEIR \cite{thakur2021} datasets. While this approach also leads to higher computational costs, the increasing availability of computational resources might make both approaches feasible. 

\subsection{Training  Adaptation}
However, Instead of directly altering the models to improve their zero-shot performance, we can use multi-task, domain-invariant or parameter-efficient learning to adapt the model training in the dense tIR pipeline.

Multi-task learning involves jointly training self-supervised tasks, dense retrieval, and extractive question answering \cite{fun2021efficient}. This approach can significantly improve the zero-shot setting and enhance the model's generalisation capabilities \cite{aghajanyan2021muppet}.

The joint training across multiple tasks mitigates the model's propensity to overfit a specific task. The multi-task learning paradigm improves performance, even when a substantial amount of labelled training data from the target domain is available \cite{maillard2021multitask}. Several approaches have been proposed within this framework: One suggests fine-tuning a universal retriever on multiple, diverse tasks, effectively training the model to handle various domains and tasks simultaneously. Another task-specific retriever ensembling approach entails training separate retrievers for distinct tasks and aggregating their predictions. As demonstrated by Li et al. \cite{li-etal-2021-multi-task-dense}, this is particularly advantageous when dealing with document corpora that display inconsistencies. To further expedite domain adaptation, combining multi-task learning with meta-learning techniques has shown promise \cite{finn2017modelagnostic, qian-yu-2019-domain, dai-etal-2020-learning, park2021unsupervised}.

Domain-invariant learning is another approach to adapting the model training in dense tIR pipelines. It aims to equip models with representations that remain invariant across different domains to enable better generalisation from a source domain to a target domain \cite{ganin2015unsupervised, ganin2016domainadversarial}. The core training objective in domain-invariant learning consists of two components, represented as follows:
\begin{equation}
L = L_{rank} + \lambda L_{DiL}
\end{equation}
where $\lambda$ is a tunable hyperparameter, $L_{rank}$ is the ranking loss typically applied to out-of-domain trainings data:
\begin{equation}
- \min_{R} \mathbb{E}_{q, d^+, d_{1\sim n}^- \in Q_o \times D_o} \mathcal{O} (R, q, d^+ , d^-_{1 \sim n})
\end{equation}
The second loss function component, $L_{DiL}$, is the domain-invariant learning loss, designed to minimise the distance between the embedding spaces of the source and target domain:
\begin{equation*}
\begin{aligned}
L_{DiL} = \min_{E_1, E_2} \mathbb{E}_{q_t \in Q_t, d_t \in D_t, q_0 \in Q_o, d_o \in D_o} \\
D (E1(q_o) || E1(q_t)) + D (E2(d_o) || E2(d_t))
\end{aligned}
\end{equation*}
where $D$ is a discrepancy function, $E_1$ corresponds to the query encoder, and $E_2$ is the document encoder. Two popular choices for the discrepancy function, $D$, have emerged. The first choice, Maximum-Mean Discrepancy (MMD), quantifies the distance between distributions in probability space. It utilises the norm of the difference between the first moments of both domain embedding distributions \cite{Quinonero2009, tzeng2014deep, long2015learning}. 
\begin{equation}
D_{MMD}(P,Q)=|| \mu_P-\mu_Q||_H
\end{equation}
where $P$ and $Q$ are the distributions of two different domains and, $\mu$ is the distribution mean, estimated by the empirical mean over batches \cite{tran2019} and $H$ is the corresponding reproducing Kernel Himert space \cite{Berlinet2004}. Additional regularisation techniques, such as minimising second moments, have been applied to enhance domain invariance \cite{Chen_2020, liu2022collaborative}. However, only focusing on the distributions' first and second moments does not guarantee their domain invariance. Another approach randomly projects both embedding probability distributions, making them one-dimensional and Gaussian-like. Afterwards, minimising the first two moments makes the whole distributions similar to one another \cite{tran2019}.

The second choice for the discrepancy function $D$ is adversarial learning. It uses a discriminator $f$ to estimate the density ratio of two distributions \cite{goodfellow2014generative}:
\begin{equation}
\min_{f} \mathbb{E}_{x \sim P/Q}-logf(l_x|x)
\end{equation}
where $l_x$ is the label indicating from which distribution ($P$ or $Q$) $x$ stems. If $f$'s training objective is modified, it can be utilised to estimate distance metrics between two distributions, like the Jesson-Shannon or Wasserstein divergence \cite{Goodfellow2014, pmlr-v70-arjovsky17a}:
\begin{equation}
D_{adv}(P,Q) = \mathbb{E}_{x \sim P/Q} L (U|f(l_x|x))
\end{equation}
where $U$ is the uniform distribution, and $L$ is either a cross-entropy or a KL divergence loss function.
The model is trained iteratively, optimising the first and second functions from above to make it impossible for $f$ to distinguish from which domain encoded embeddings stem since both domains share the same distribution space.
Adversarial learning methods have been applied in various natural language processing tasks. Lee et al. \cite{lee2019domainagnostic} show that adversarial learning can be used for question-answering tasks and performs better than a baseline model. Wang et al. \cite{wang2019adversarial} apply adversarial learning to synthetically generated queries. Shrivastava et al. \cite{shrivastava2022qagan} introduce the annealing trick, which prevents the discriminator from providing misleading signals during early training stages. Xin et al. \cite{xin-etal-2022-zero} apply domain invariant learning to dense retriever training. They propose maintaining a momentum queue averaging over past batches, so feeding all data points to the classifier at every batch is not needed. The authors aim to balance the efficiency and robustness of the domain classifier and show that the performance is better than that of a baseline model. However, Xin et al. do not compare domain invariant learning to other domain adaptation techniques.

Balancing the ranking and domain-invariant learning loss requires careful parameter tuning. When adversarial learning is employed, adjusting the update rates of the dense retriever and discriminator becomes even more challenging \cite{wang2020crossdomain, shrivastava2022qagan}. The training process for domain-invariant learning is generally more complex than other approaches. While domain-adaptive training is a promising approach, its potential when combined with other techniques or its practicality in low-resource settings must be explored further \cite{shen2022lowresource}.

Parameter-efficient learning is the third approach to improving the zero-shot performance by adapting the model training in the tIR pipeline. It is based on modifying only a subset of the model's parameters, allowing it to sustain general knowledge learnt in pre-training and thus better generalisation in low-resource setup. The different pre-training strategies this approach can enhance are task-adaptive pre-training, which mimics the retrieval task in a self-supervised way \cite{lee2019latent, chang2020pretraining, ram2022learning}. Retrieval augmented pre-training which aims to improve the modelling capacity of pre-trained language models by integrating a so-called knowledge retriever that provides referring context for the Masked Language Modelling task \cite{guu2020realm} and representation enhanced pre-training, which aims to enhance the amount of information stored in the embedding of the "[CLS]" token. Representation enhanced pre-training is performed by autoencoder enhanced pre-training \cite{Gao2021b, wang2021, gao2021unsupervised, lu-etal-2021-less, wang2023simlm, xiao2022retromae, liu2022masked} or contrastive learning enhanced pre-training \cite{Gao2021, yan2021consert, xu2022laprador, Ma_2022}. All these pre-training techniques can aid domain adaptation of dense retrievers in low-resource setups in varying ways. 
Possible approaches to applying parameter-efficient learning - to sustain the general knowledge a model acquires during pre-training - are: adapters \cite{houlsby2019parameterefficient} or prompt-tuning \cite{li2021prefixtuning, liu2021gpt}. Tam et al. \cite{tam2022parameterefficient} even use parameter-efficient prompt tuning for retrieval tasks in-domain, cross-domain and cross-topic while only fine-tuning 0.1\% of the model's parameters, leading to better generalisation performance than in the case of fine-tuning all parameters.

\subsection{Ranking Adaptation}
Ranking adaptation of dense retrievers is performed by integrating sparse-retrieval to combine the strengths of both approaches. Either the dense and sparse retriever relevance scores are directly combined \cite{Wang2021bm25, xu2022laprador, karpukhin-etal-2020-dense} or an unsupervised retriever is improved by interpolating BM25 and DR relevance scores \cite{ram2022learning, ren2023thorough}. Another approach utilises RM3 \cite{Jaleel2004UMassAT} built on top of the sparse results for selecting deep retriever results as the final list \cite{kuzi2020leveraging} and Chen et al. \cite{Chen_2022} use robust rank-based scores leading to significant zero-shot performance improvements.
Another option is to use dense retrievers to generate term weights for improving sparse retrieval results \cite{Dai2020} or to apply a classifier to decide whether to use dense, sparse or hybrid retrieval results for each query \cite{arabzadeh2021predicting}. 
However, while integrating sparse retrieval into dense retrieval systems enhances their term-matching capabilities, this approach requires maintaining two indexing systems simultaneously, which is often not feasible in practice. Nevertheless, some approaches that do not require a dual index but still enhance the term-matching capacities of dense retrieval systems exist.
It is, for example, possible to learn low-dimensional dense lexical representations \cite{lin2023aggretriever, lin2021densifying} or to train a dense lexical retriever based on weakly supervised data constructed by BM25 \cite{chen2022salient}. Another option are pre-trained language models which have lexical awareness (like SPLADE \cite{formal2021splade}) \cite{zhang2023led}. They are trained using lexicon-augmented contrastive learning and rank-consistent regularisation. 

\subsection{Discussion} \label{discussion domain adaptation}
Apart from parameter-efficient learning, all the introduced methods lead to higher computational costs for building dense tIR systems while improving their zero-shot capabilities. They also require different data inputs, as shown in \Cref{tab:da}.

\begin{table}[ht] 
  \centering
  \caption{Domain-Adaptation Approaches}
  \label{tab:da} 
  \begin{adjustbox}{width=\columnwidth,center}
  \begin{tabular}{@{\extracolsep{5pt}}|c|c|c|} 
    \hline
    Adaptation Level & Method & Data Requirements Target Domain \\
    \hline
    \SetCell[r=3]{c}Data 
    & Query Generation & Only Text \\
    & Contrastive Learning & Only Text \\
    & Knowledge Distillation & Text and Queries \\
    \hline
    \SetCell[r=2]{c}Model
    & Size & Labelled Data \\
    & Capability & Labelled Data \\
    \hline
    \SetCell[r=3]{c} Training 
    & Multi-task & Labelled Data \\
    & Domain-Invariant & Labelled Data \\
    & Parameter-Efficient & Labelled Data \\
    \hline
    \SetCell[r=1]{c}Ranking 
    & Integrating Sparse Retrieval & Labelled Data \\
    \hline
  \end{tabular}
  \end{adjustbox}
\end{table}

Theoretically, all methods listed in \Cref{tab:da} can be combined. Nevertheless, combining methods further increases the computational costs of training the system. Also, utilising contrastive learning and query generation in the same tIR system does not make sense since they have the same objective of generating (pseudo) query-document pairs. Thus, adding one or the other is sufficient. However, combining knowledge distillation with query generation or contrastive learning can generate fine-grained (pseudo) relevance judgments for the generated (pseudo) query document pairs, mitigating the requirement of independent queries and documents for knowledge distillation.

Ren et al. \cite{ren2023thorough} empirically compared different data-level approaches to adapt dense retrievers to a target domain in a low-resource setting. They found that LaPraDoR \cite{xu2022laprador} outperformed other methods. It combines contrastive learning with the integration of sparse retrieval. However, if the BM25 part is removed, the performance on target domain sets with considerable vocabulary overlap drops. Thus, integrating lexical matching can improve retrieval performance on datasets with high vocabulary overlap. For target data with low vocabulary overlap, the improvement was lower. 
GPL \cite{wang2022} achieved competitive performance since it improved vocabulary overlap and query type distribution similarity between target and source data. It also increases the available (pseudo) labelled data through query generation. 

\section {Multilingual Semantic Search} \label{multilingual}
Since most available labelled datasets are in English, developing multilingual semantic search systems or even monolingual dense retrievers for low-resource languages remains challenging. Two exceptions are mMARCO \cite{Bonifacio2021} - a multilanguage version of MS MARCO, comprised of 13 languages, created by machine translation - and Mr. TYDI \cite{zhang2021mr} - comprised of 11 languages. 

Multilingual dense retrievers are trained on monolingual text representations from multiple languages, generating multilingual text representations that enable a single model to perform in multiple languages \cite{lange-etal-2020-choice, Devlin2018, Conneau2019}. In zero-shot settings, they can be fine-tuned on the target task in a high-resource language and then applied to unseen target languages. Their performance is improved if we add a small amount of target task and target language data. This transfer between languages is achieved through creating a shared multilingual embedding space by mapping between the different embedding spaces of the two languages, resulting in similar feature vectors for embeddings in the two vectors spaces \cite{lauscher-etal-2020-zero, hedderich-etal-2020-transfer, mikolov2013exploiting, joulin2018loss}. With this method, single languages can be differently embedded by the same model and mapped to the same vector space afterwards \cite{cao2020multilingual}. 

Recently, Feng et al. \cite{feng-etal-2022-language} introduced LaBSE, which creates sentence embeddings with 768 dimensions and trains 471M parameters. LaBSE performs well on sentence alignment tasks, partly attributed to the multilingual model pre-training. However, the high dimensional embeddings and the many parameters make adapting LaBSE to downstream tasks computationally expensive. As an alternative, LEALLA \cite{mao2023lealla} is a distilled encoder model to create multilingual sentence embeddings with lower dimensions, reducing the computational overhead for down-stream tasks as well as creating the embeddings themselves while sustaining competitive performance on benchmark datasets for parallel sentence alignment. Before LEALLA, no multilingual pre-trained lightweight language models were available. Thus, for training LEALLA, the authors used LaBSE as a teacher in knowledge distillation, combining feature and logit distillation.

In contrast to these two approaches, LASER \cite{Artetxe_2019} trains an encoder-decoder LSTM model utilising a translation task. This translation-based training enables LASER to perform well on exact translation tasks. Nevertheless, LASER's ability to accurately recognise the similarity of sentences that are not exact translations could be improved. Another approach is the Multilingual Universal Sentence Encoder (mUSE) \cite{chidambaram-etal-2019-learning, yang-etal-2020-multilingual}. MUSE was trained on general domain data in multiple languages, utilising a multi-task setup, namely a translation ranking task in which, given several incorrect and one correct translation, the latter needs to be identified. However, this task requires hard negatives to enable state-of-the-art performance, leading to a significant computational overhead \cite{guo-etal-2018-effective}.
Reimers and Gurevych \cite{reimers-gurevych-2020-making} further explore the application of knowledge distillation to make monolingual pre-trained language models multilingual.

Another application of multilingual language models is cross-lingual retrieval. For example, XOR QA \cite{asai-etal-2021-xor} answers a question in one language based on a document in another language, and Asai et al. \cite{asai2021one} utilise iterative training to fine-tune a multilingual PLM to build a cross-lingual dense passage retriever which can retrieve candidate passages from a multilingual document collection.

\section{Discussion} \label{discussion}
An interesting approach to explore further would be utilising cross-lingual models for query generation on texts from a specialised target domain. Would they outperform existing machine translation systems? Furthermore, the approach to use a capable, multilingual teacher in knowledge distillation, as already investigated by Mao et al. \cite{mao2023lealla} as well as Reimers and Gurevych \cite{reimers-gurevych-2020-making} might be expandable to performing domain-adaptation if a pre-trained student model utilising parameter efficient learning is applied. Moreover, domain-invariant learning and the training of multilingual models share a similar objective. Therefore, it would be interesting to see whether issues arising from the complex loss formulation for domain-invariant learning could be mitigated by applying a mapping function to create a joint space for source- and target-domain embeddings. Additionally, would it be possible to extend a shared source- and target domain vector space to include multilingual embeddings? 

\section{Outlook and Conclusion} \label{outlook}
This literature review explored how to perform domain adaptation for multilingual semantic search. We introduced a systematic clustering of domain adaptation methods in a low resource setup, based on parts of the tIR system they alter, summarised in \Cref{tab:da}. Focusing on tIR system parts and data requirements of the different approaches helps clarify their combination possibilities. Afterwards, we gave an overview of recent advances in multilingual semantic search, followed by a discussion on combining it with the introduced domain adaptation approaches. The domain adaptation approaches we introduced focus on monolingual semantic search. However, some multilingual models already adapted knowledge distillation \cite{mao2023lealla, reimers-gurevych-2020-making}. Combining other domain adaptation techniques with multilingual semantic search is an exciting research direction to investigate further. The questions we raised in \Cref{discussion} might spark some ideas in that regard.
A limitation of this work is that we did not distinguish sentence encoders from other sequence encoding strategies in our analysis. Doing so in future work would be fruitful. 

\addcontentsline{toc}{section}{References}
\bibliographystyle{ACM-Reference-Format}
\setlength{\bibsep}{0.4\baselineskip}
\bibliography{Thesis}

\appendix
\section{History of textual information retrieval} \label{history}
The textual information retrieval task was theorised as early as 1950 by T. Joyce and R. M. Needham \cite{Joyce1958}. They asked how texts can be indexed and picked representative terms to retrieve relevant information. 
Following that idea, the "bag-of-words" assumption led to a vector space model that encoded the query and the document in sparse term-based vectors \cite{Salton1962, Salton1975}. Many different approaches to constructing these sparse vectors exist. A first approach, called tf-idf, allows for a relevance measure based on the text vector's and query vector's lexical similarity \cite{Salton1988, Aizawa2003, Robertson2004}. As a second approach, inverted indexing utilises the same representation idea as tf-idf but makes text retrieval more efficient by organising documents in term-oriented posting lists \cite{Zobel2006, Zobel1998}. Probabilistic relevance frameworks are a third approach to solving the text retrieval task and modelling the relevance of documents. They enable a more in-depth understanding of the retrieval mechanisms. One example of such a framework is BM25 \cite{Robertson1994, Robertson2009}, but multiple more (statistical) language modelling approaches exist \cite{Zhai2009}. \newline
The rise of machine learning approaches led to the use of learning to rank algorithms for text retrieval tasks \cite{Liu2011, Li2011}. These algorithms apply supervised machine learning models to rank the relevance of retrieved documents but require human feature engineering. \newline
The availability of more computational power and the application of stochastic gradient descent for neural networks resulted in neural information retrieval, a deep learning approach to text retrieval tasks. This approach no longer required hand-designed feature engineering \cite{Huang2013, GuoetAl2017, Mitra2017, Guo2019}. Neural information retrieval models map the query and the text corpus into low-dimensional vectors in latent representation space. Thus, they encode the textual input in dense vectors. This process is also called embedding. The similarity between the query and the document vector estimates the relevance of a text for a query. Thus, the relevance of information is no longer estimated by lexical but by semantic similarity.
Contrary to this approach, the sparse vectors constructed by classical vector space models encode explicit term dimensions. In contrast, the dense vectors constructed by neural IR models encode the latent semantic characteristics of language. The so-called transformer architecture was the subsequent historical development in the textual information retrieval task after neural IR. Nowadays, transformers are the basis for most NLP tasks. They were initially proposed to model any sequence data by utilising the self-attention mechanism through which every token attends to every other token in the sequence \cite{Vaswani2017, Hochreiter1997LSTM, Cho2014}. Their two significant advantages are that they can be trained in parallel and scaled up easily. The use of transformers for NLP tasks, as well as the availability of large-scale labelled retrieval datasets like MS MARCO\cite{Bajaj2016} gave rise to pre-trained language models \cite{Devlin2018, Liu2019, Brown2020, fan2022pretraining}. The pre-trained language models use different self-supervised loss functions and are trained on large-scale general document corpora. Fine-tuning the models enables their transfer to downstream tasks \cite{Liu2021, zhao2022dense}. They facilitated a broader representation of semantics and language in general. The most common pre-trained language model is called BERT \cite{Devlin2018}. It employs a deep bi-directional architecture and word masking to improve text encoding into dense vectors. BERT can encode the general English language.
The recent developments after BERT can be categorised into three different research fields. Firstly, different pre-training approaches are explored \cite{Liu2019}. Secondly, the bi-directional representation is refined \cite{Ma2019}, and thirdly, a compressed and thus more lightweight version of BERT was developed, called distilBERT \cite{Sanh2019}. Nowadays, pre-trained dense retrievers are the "gold standard" for solving tIR tasks due to their astonishing representation of documents and ability to answer more complex queries \cite{zhao2022dense}. However, most pre-trained dense retrievers were trained on English data containing general, informal knowledge. Transforming them to other languages or specialised domains requires further fine-tuning on labelled data in the desired domain and language. Such data, as mentioned in \Cref{introduction}, is rarely available.

\end{document}